\documentclass[12pt]{article}
\usepackage[xdvi]{graphicx}
\usepackage{amsfonts}
\begin{document}
\newcommand{\todo}[1]{{\em \small {#1}}\marginpar{$\Longleftarrow$}}
\newcommand{\labell}[1]{\label{#1}\qquad_{#1}} %{\label{#1}} %

\rightline{UPR-1037-T, DCPT-03/19} \rightline{hep-th/0304150} \vskip 1cm

%%          Title here
%%

\begin{center}
{\Large \bf Holography beyond the horizon and cosmic censorship}
\end{center}
\vskip 1cm

\renewcommand{\thefootnote}{\fnsymbol{footnote}}
\centerline{\bf
Thomas S. Levi${}^1$\footnote{tslevi@student.physics.upenn.edu} and Simon
F. Ross${}^2$\footnote{S.F.Ross@durham.ac.uk}}
\vskip .5cm
 \centerline{${}^1$\it David Rittenhouse Laboratories, University of
Pennsylvania,} \centerline{\it Philadelphia, PA 19104, USA}
\vskip .5cm \centerline{ ${}^2$\it Centre for Particle Theory, Department of
Mathematical Sciences} \centerline{\it University of Durham, South Road, Durham DH1 3LE, U.K.}

\setcounter{footnote}{0}
\renewcommand{\thefootnote}{\arabic{footnote}}

%%          Text starts here
%%

\begin{abstract}
We investigate the description of the region behind the event horizon
in rotating black holes in the AdS/CFT correspondence, using the
rotating BTZ black hole as a concrete example. We extend a technique
introduced by Kraus, Ooguri and Shenker~\cite{kos}, based on
analytically continuing amplitudes defined in a Euclidean space, to
include rotation. In the rotating case, boundary amplitudes again have
two different bulk descriptions, involving either integration only
over the regions outside the black holes' event horizon, or
integration over this region and the region between the event horizon
and the Cauchy horizon (inner horizon). We argue that generally, the
holographic map will relate the field theory to the region bounded by
the Cauchy horizons in spacetime. We also argue that these results
suggest that the holographic description of black holes will satisfy
strong cosmic censorship.
\end{abstract}

\section{Introduction}

Black holes play a central role in the AdS/CFT correspondence. The
original proposal of a large $N$ duality~\cite{dual1} was largely motivated
by the successes of string theory in explaining the black hole
entropy, and soon afterwards, it was realised~\cite{dualbh} that the black
hole geometries are directly related to thermal states in the gauge
theory. This led to non-trivial predictions for the phase structure of
the strongly-coupled gauge theory, and placed our understanding of
string theory's successes in accounting for the black hole entropy on
a much firmer footing. However, our understanding of the description
of the black hole {\it geometry} (as opposed to its thermodynamic
properties) from the dual field theory point of view remains very
weak. It has proved difficult to address even the simplest questions
about observations in a black hole spacetime from the field theory
point of view. For example, we do not know how the `one-way' nature of
the event horizon is (approximately) enforced in the field theory, nor
do we know how to describe the observations of an observer falling
across the black hole horizon.

These difficulties reflect broader problems in understanding the
Lorentzian aspects of the AdS/CFT correspondence. The connection
between thermodynamics of black holes and thermodynamics of gauge
theory can be understood using the Euclidean black hole solutions. But
in this Euclidean section, there is no region `behind the event
horizon': in positive definite signature there is in general no
coordinate-independent notion of an event horizon. The usual analytic
continuation back to Lorentzian signature maps the Euclidean section
onto the region outside the event horizon of an eternal black hole. It
has sometimes been suggested that this implies that the dual field
theory describes only physics in the region outside the event
horizon. However, as argued in~\cite{resol}, this seems very unlikely:
there is a Cauchy surface for the black hole spacetime which lies
entirely outside the event horizon. Thus, if the dual field theory
provides a complete description of the physics in this region, it will
also determine what is happening behind the horizon. That is, degrees
of freedom which pass behind the event horizon should still be encoded
in the dual field theory in some way, which could involve some form of
black hole complementarity~\cite{compa,compb}.

Attempts to explore this encoding were made
in~\cite{holog,geon,eternal,hub}. In~\cite{holog,geon}, attention was
focused on the description of black holes formed from collapse and on
the geon, so the geometry has a single asymptotic boundary, and the
full geometry including the region behind the horizon should be
described by some appropriate state in the dual field theory. One
obstacle to progress in these cases was that the correct state in the
field theory is not easily identified. In~\cite{eternal}, the eternal
BTZ black hole was considered, and a dual description of the geometry
in terms of a specific entangled state of the two CFTs living on the
two asymptotic boundaries of the spacetime was proposed (building on
previous work~\cite{thermof,orb1,orb2,hads}). This provides a concrete
context in which one can ask what information about the region behind
the event horizon is contained in correlation functions of CFT
operators in this entangled state. It was argued in~\cite{eternal}
that correlation functions involving operators in both CFTs would
indeed probe the region behind the horizon.

In~\cite{kos}, substantial progress in implementing these ideas was
achieved by introducing two key ingredients: first, the calculation of
higher $n$-point correlators using bulk-boundary propagators was
considered. This involves an interaction vertex which is integrated
over the bulk, allowing the explicit identification of contributions
from different regions of the spacetime. Second, a novel analytic
continuation was found, which relates the integration of the bulk
interaction vertex over the Euclidean black hole to an integration
over the whole black hole spacetime, including the regions inside the
black hole event horizon. This amplitude could also be analytically
continued in the standard way, to obtain a representation in terms of
an integral just over the region outside the black hole event horizon.
(These alternative analytic continuations were also discussed
in~\cite{berk}.)  This technique thus provides an explicit realisation
of the ideas of black hole complementarity~\cite{compa,compb}.  It
shows that there is a representation in which the field theory treats
the black hole singularity, and not the event horizon, as the end of
the classical spacetime description.

In this paper, we will extend the analysis of~\cite{kos} to consider
the rotating BTZ black hole.  This involves a generalisation of the
analytic continuation described in~\cite{kos}. We will see that the field theory
calculation can only be related to an integral over the region outside
the inner horizon of the black hole; this suggests the
region inside the inner horizon, where the singularities are visible,
is unphysical from the field theory point of view.

We proceed as in~\cite{kos}, investigating the definition of
correlation functions in the entangled CFTs on the two asymptotic
boundaries of the rotating black hole by analytic continuation from a
Euclidean solution. In the bulk description, a boundary correlation
function is calculated by evaluating appropriate Feynman diagrams
constructed from bulk-boundary and bulk-bulk propagators, with the
interaction vertices integrated over the bulk spacetime. As we
analytically continue the boundary points from the Euclidean to the
Lorentzian section, we must deform the contour of integration for
these interaction vertices in some way.

We will investigate two different ways of deforming this contour,
which lead to different representations of the same Lorentzian
correlation function. In the first case, we deform in the usual way,
taking the integration over the Euclidean spacetime to be an
integration over the regions outside the black hole event horizons. We
then find a new `Kruskal-like' coordinate system, which allows us to
define a deformation of the contour as we analytically continue which
can be interpreted as giving an integration of the interaction
vertices over the whole region between the past and future Cauchy
horizons (inner horizons), including both the regions between the
inner and outer horizons and the exterior asymptotic regions. Thus, in
this case, the boundary correlation function can be related to an
integral over the Cauchy development of a Cauchy surface connected to
the asymptotic boundaries. There appears to be no representation in
which the computation of boundary observables involves the spacetime
region beyond the Cauchy horizon. The field theory seems to see the
Cauchy horizon (and not the event horizon) as a natural boundary. This
supports the argument given previously, that the region inside the
event horizons must be described by the field theory because it is
included in this Cauchy development.

There is another reason for expecting the spacetime to be cut off at
the Cauchy horizon: an observer who crosses it will see the naked
timelike singularity inside the black hole, leading to a violation of
strong cosmic censorship~\cite{scc}. The fact that the CFT only
appears to describe the region outside the Cauchy horizon may
therefore be interpreted as an interesting manifestation of cosmic
censorship in the AdS/CFT correspondence.  It has previously been
argued that violations of strong cosmic censorship in such charged
black hole spacetimes are prevented by an instability of the Cauchy
horizon: it is a surface of infinite blueshift~\cite{blues}, and is
therefore unstable to generic perturbations of the exterior
geometry~\cite{instab}.  It is not clear if the fact that the CFT sees
the Cauchy horizon as the end of the classical spacetime in rotating
black holes is a reflection of this instability, or represents a truly
independent mechanism for enforcing cosmic censorship. An important
direction for future work is to look for signs of singular behaviour
in the CFT associated with the region of spacetime near Cauchy
horizons, and to understand the conjectured breakdown of the classical
spacetime picture there.

In section~\ref{rot}, we review the rotating BTZ black hole solution,
and discuss the analytic continuation relating the Euclidean and
Lorentzian solutions in the standard BTZ coordinate
system~\cite{btz}. This analytic continuation was also recently
discussed in~\cite{eskos}, where the propagation of strings in this
geometry was considered. In section~\ref{disc}, we introduce a new
coordinate system, in which the analytic continuation relates the
Euclidean section to the regions outside the event horizon and between
the event horizon and the Cauchy horizon. Our discussion follows that
in~\cite{kos} very closely. Section~\ref{concl} contains some
concluding remarks and thoughts on future directions.

\section{Rotating BTZ black holes}
\label{rot}

We will begin by reviewing the relevant features of the BTZ solution,
the propagators for scalar fields on this spacetime, and the analytic
continuation to define a related spacetime with positive definite
metric.  In static coordinates, the rotating BTZ solution
of~\cite{btz} is (we are setting the AdS length scale associated with the
cosmological constant to one)
\begin{equation} \label{rbtz}
ds^2 = - {(r^2- r_+^2)(r^2-r_-^2) \over r^2} dt^2 + {r^2 dr^2 \over
  (r^2 - r_+^2)(r^2 - r_-^2)} + r^2 \left( -{r_+ r_- \over r^2} dt +
  d\bar \phi \right)^2.
\end{equation}
The parameters $r_\pm$ are related to the mass and angular momentum of
the solution through
\begin{equation}
r_\pm^2 = {M \over 2} \left[ 1 \pm \left( 1 - {J^2 \over M^2}
  \right)^{1/2} \right],
\end{equation}
so $M = r_+^2 + r_-^2$, $J = 2 r_+ r_-$. This solution is locally
AdS$_3$; this is most easily seen by relating the coordinate system
used in (\ref{rbtz}) to the embedding coordinates $(T_1,T_2,X_1,X_2)$
for the AdS$_3$ hyperboloid, which satisfy $T_1^2 + T_2^2 -X_1^2
-X_2^2 = 1$. They are related by
\begin{equation}
T_1 = \sqrt{\alpha} \cosh( r_+ \bar \phi - r_- t),
\end{equation}
\begin{equation}
T_2 = \sqrt{\alpha-1} \sinh( r_+ t - r_- \bar \phi),
\end{equation}
\begin{equation}
X_1 = \sqrt{\alpha} \sinh( r_+ \bar \phi - r_- t),
\end{equation}
\begin{equation}
X_2 = \sqrt{\alpha-1} \cosh( r_+ t - r_- \bar \phi),
\end{equation}
where $\alpha = {r^2-r_-^2 \over r_+^2 - r_-^2}$.  In discussing
physics near the horizon, it is often useful to introduce a different
angular coordinate, $\phi = \bar \phi - r_- t/r_+$. In terms of this
coordinate,
\begin{equation} \label{rbtzb}
ds^2 = - {(r^2- r_+^2)(r^2-r_-^2) \over r^2} dt^2 + {r^2 dr^2 \over
  (r^2 - r_+^2)(r^2 - r_-^2)} + r^2 \left[ {r_- \over r_+ r^2} (r^2 -
  r_+^2) dt + d \phi \right]^2,
\end{equation}
and the relation to the embedding coordinates becomes
\begin{equation}
T_1 = \sqrt{\alpha} \cosh( r_+ \phi),
\end{equation}
\begin{equation}
T_2 = \sqrt{\alpha-1} \sinh\left( {r_+^2 - r_-^2 \over r_+} t - r_-
\phi \right),
\end{equation}
\begin{equation}
X_1 = \sqrt{\alpha} \sinh( r_+ \phi),
\end{equation}
\begin{equation}
X_2 = \sqrt{\alpha-1} \cosh\left( {r_+^2 - r_-^2 \over r_+} t - r_-
\phi \right).
\end{equation}
This change of coordinates gives a metric where the $g_{t \phi}$ cross
term vanishes at the horizon, so $\phi$ is a co-rotating coordinate
near the horizon (the original coordinate system is adapted to
discussing infinity; the $g_{t \bar \phi}$ cross term vanishes at
infinity).

\begin{figure}
\begin{center}
    \includegraphics[width=0.8\textwidth]{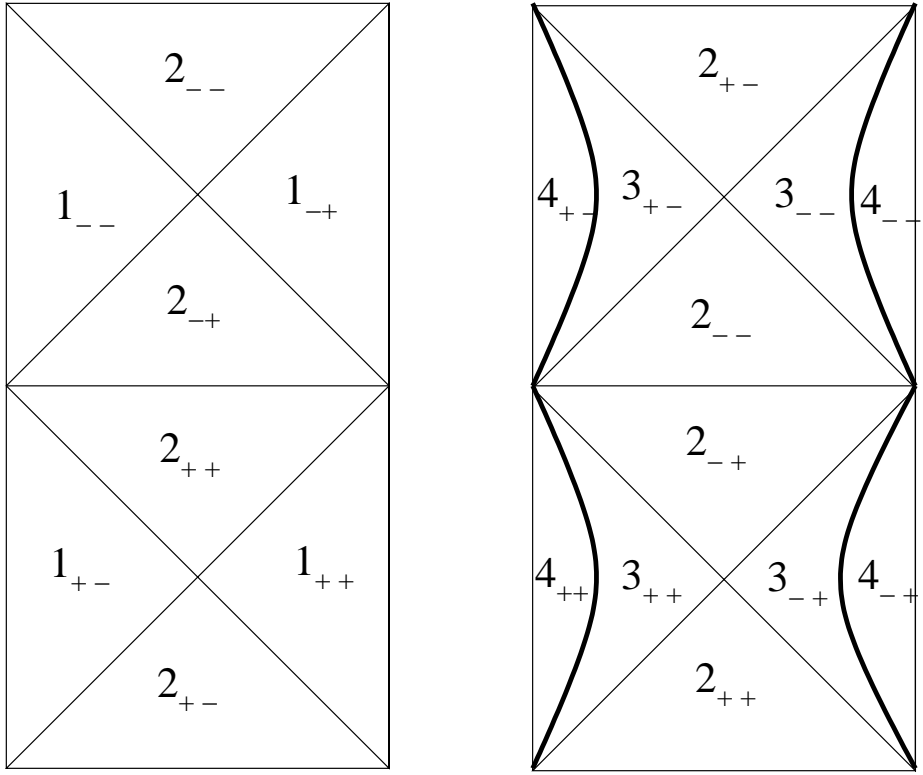}
\end{center}
\caption{Slices through AdS$_3$ in global coordinates, showing the
  regions of AdS$_3$ determined by the norm of the Killing vector $\xi
  =\partial_\phi$ in the coordinates (\ref{rbtzb}). In region
  1, $\xi \cdot \xi \geq r_+^2$; in region 2, $r_+^2 \geq \xi \cdot \xi \geq
  r_-^2$; in region 3, $r_-^2 \geq \xi \cdot \xi > 0$; in region 4, $\xi
  \cdot \xi \leq 0$.} \label{fig:pen}
\end{figure}

This spacetime is naturally divided into several regions, determined
by the norm of the Killing vector $\xi = \partial_\phi$ along which we
identify. If $\xi \cdot \xi \geq r_+^2$, we are outside the black hole's
event horizon; this is region 1. If $r_+^2 \geq \xi \cdot \xi \geq r_-^2$,
we are between the event horizon and the Cauchy horizon; this is
region 2. If $r_-^2 \geq \xi \cdot \xi > 0$, we are between the Cauchy
horizon and the `singularity' of the black hole; this is region
3. Finally, in the global AdS$_3$ space before making identifications,
there is a region where $\xi \cdot \xi \leq 0$, which we call region
4---this region is omitted from the black hole spacetime. Because we
cut out region 4, treating the surface $\xi \cdot \xi = 0$ as a
singularity of the rotating black hole spacetime, this black hole
solution has no closed timelike curves. The division of global AdS$_3$
into these regions is depicted in figure~\ref{fig:pen}.

As in~\cite{kos}, it will be convenient later to note that the
division into regions can also be described in terms of the embedding
coordinates. Since $T_1^2 - X_1^2 = \alpha$, and $T_2^2 - X_2^2 =
1-\alpha$, the regions can also be distinguished by
\begin{eqnarray}
\mbox{ region 1:} & T_1^2 - X_1^2 \geq 0, &  T_2^2 - X_2^2 \leq 0 \\
\mbox{ region 2:} & T_1^2 - X_1^2 \geq 0, & T_2^2 - X_2^2 \geq 0  \nonumber \\
\mbox{ regions 3 \& 4:} & T_1^2 - X_1^2 \leq 0, & T_2^2 - X_2^2 \geq 0
\nonumber
\end{eqnarray}
We also note that each of the regions consists of several disconnected
components; these are conveniently distinguished by the sign of
certain combinations of the embedding coordinates. We adopt the same
procedure as~\cite{kos}, denoting the different regions by $A_{\eta_1
  \eta_2}$, where $A=1,\ldots,4$ and $\eta_1, \eta_2 = \pm$ are the
signs of the two combinations $T_1 + X_1$ and $T_2 + X_2$. Thus, for
example, the ordinary region outside the black hole covered by the
coordinates in (\ref{rbtz}), where $T_1 + X_1 = \sqrt{\alpha} e^{r_+
  \phi} >0 $ and $T_2 + X_2 = \sqrt{\alpha-1} \exp\left({ r_+^2 -
  r_-^2 \over r_+} t - r_- \phi\right)>0 $, is denoted $1_{++}$.

We want to investigate the calculation of field theory amplitudes
using the bulk-boundary propagators for a scalar field on this
solution, defining the propagators by analytic continuation from a
Euclidean section. The rotating BTZ metric has no Euclidean
continuation {\it per se}; we cannot define a real Euclidean metric by
complexifying the coordinates in the above metric. However, there is a
standard analytic continuation to a Euclidean space defined by
analytically continuing simultaneously in the coordinates and in the
parameter $r_-$.\footnote{Of course, global AdS$_3$ does have a
Euclidean section. If we think of the BTZ black hole as a quotient of
global AdS$_3$, this analytic continuation of $r_-$ amounts to
redefining the identification we consider as we perform the analytic
continuation from the Lorentzian to the Euclidean section in AdS$_3$.}
We set $t = -i\tau$, $r_- = -i \tilde r_-$ (so $\tilde J = i J = 2 r_+
\tilde r_-$). With this analytic continuation, the metric
(\ref{rbtzb}) becomes
\begin{equation} \label{erbtzb}
ds^2 = {(r^2- r_+^2)(r^2+ \tilde r_-^2) \over r^2} d\tau^2 + {r^2 dr^2
  \over (r^2 - r_+^2)(r^2 + \tilde r_-^2)} + r^2 \left[ -{\tilde r_-
  \over r_+ r^2} (r^2 - r_+^2) d\tau + d \phi \right]^2.
\end{equation}
At $r=r_+$, there is a conical singularity in this metric, as $g_{\tau
\tau} \to 0$. We must therefore identify $\tau$ periodically with
period $\Delta \tau = 2\pi r_+/(r_+^2 + \tilde r_-^2)$. In this
coordinate system, the Euclidean metric (\ref{erbtzb}) is smooth with
the separate identifications $\phi \sim \phi + 2\pi n$ and $\tau \sim
\tau + m \Delta \tau$, with $m,n$ integers. One normally expects a
rotating black hole to involve a twisted identification $\bar \phi
\sim \bar \phi + \Omega \beta$, $\tau \sim \tau + \beta$, where
$\beta$ and $\Omega$ are then interpreted as temperature and a
rotational chemical potential respectively. In the present
coordinates, this twisting is hidden in the fact that the cycles that
are orthogonal at large distances are given by $\bar \phi$ and $\tau$,
not $\phi$ and $\tau$.

Let us now consider a scalar field of mass $m$ propagating in this
geometry.  The bulk-boundary propagator in the BTZ black
hole~(\ref{rbtz}) can be obtained by the method of images from the
propagator in AdS$_3$. The result is~\cite{esko}
\begin{eqnarray}
K^{1_{++}1_{++}}(x,b') = \sum_{n=-\infty}^{\infty} && \left\{ -
\sqrt{\alpha-1} \cosh[r_+ \Delta t - r_-( \Delta \bar \phi + 2\pi n)]
\right. \\ &&+ \left. \sqrt{\alpha} \cosh[ r_+ (\Delta \bar \phi +
  2\pi n) - r_- \Delta t] \right\}^{-2h_+},
\end{eqnarray}
where $x = (t,\bar \phi,r)$ is the bulk point, $b' = (t',\bar \phi')$
is the boundary point, $\Delta t = t - t'$, $\Delta \bar \phi = \bar
\phi - \bar \phi'$, and $2 h_+ = 1 + \sqrt{1 + m^2}$. We will rewrite
this in the coordinate system~(\ref{rbtzb}), and introduce the
notation $\Delta \phi_n \equiv \Delta \phi + 2\pi n$ for convenience,
giving
\begin{equation} \label{lorp}
K^{1_{++}1_{++}}(x,b') = \sum_{n=-\infty}^{\infty} \left\{ - \sqrt{\alpha-1}
\cosh\left[{(r_+^2 - r_-^2) \over r_+}
\Delta t - r_- \Delta \phi_n) \right] + \sqrt{\alpha} \cosh( r_+
\Delta \phi_n) \right\}^{-2h_+}.
\end{equation}
In writing this form for the propagator, we have assumed that both
boundary and bulk points are in region $1_{++}$, outside the
horizon. To obtain the propagator for the bulk point in other regions
requires analytic continuation of this propagator, as explained
in~\cite{kos}. For example, the propagator for the boundary point in region
$1_{++}$ and the bulk point in region $2_{++}$ is given by
\begin{equation} \label{lorpch}
K^{2_{++}1_{++}}(x,b') = \sum_{n=-\infty}^{\infty} \left\{ - \sqrt{1-\alpha}
\sinh\left[{(r_+^2 - r_-^2) \over r_+}
\Delta t - r_- \Delta \phi_n) \right] + \sqrt{\alpha} \cosh( r_+
\Delta \phi_n) \right\}^{-2h_+}.
\end{equation}
In particular, if we consider a bulk point on the inner horizon, where
$\alpha=0$, we see that the propagator remains finite; because of the
$r_- \Delta \phi_n$ term in the $\sinh$, the summand is not
independent of $n$.

An $i\epsilon$ prescription for this propagator will be defined by
analytic continuation from the Euclidean propagator on the metric
(\ref{erbtzb}). The Euclidean propagator can be obtained by making the
replacements $t = -i \tau$, $r_- = -i \tilde r_-$ in (\ref{lorp}),
which gives
\begin{equation}
K_E(x,b') = \sum_{n=-\infty}^{\infty} \left\{ - \sqrt{\alpha-1}
\cos\left[{(r_+^2 + \tilde r_-^2) \over r_+}
\Delta \tau - \tilde r_- \Delta \phi_n) \right] + \sqrt{\alpha} \cosh( r_+
\Delta \phi_n) \right\}^{-2h_+}.
\end{equation}
One can verify by explicit calculation that this agrees with the
propagator obtained by the method of images from the propagator on
Euclidean AdS$_3$; note that in the Euclidean section $\alpha = (r^2 +
\tilde r_-^2)/(r_+^2 + \tilde r_-^2)$.

The bulk description of a generic field theory amplitude involves
connecting a truncated Feynman diagram built from bulk-bulk
propagators and vertices to the boundary points using these
bulk-boundary propagators. We want to define amplitudes by analytic
continuation: we therefore start with the boundary points on the the
boundary of the Euclidean spacetime (\ref{erbtzb}) (which is a torus),
and the bulk vertices integrated over (\ref{erbtzb}). We analytically
continue the locations $t'_i$ of the boundary points from pure
imaginary back to real values, and at the same time analytically
continue the parameter $r_-$, to obtain an amplitude in the Lorentzian
field theory. We want to see how we should deform the contour of
integration for the bulk vertices as we perform this analytic
continuation.\footnote{The discussion of analytic continuation in the
boundary field theory itself is essentially unchanged from the
non-rotating case; we will therefore not repeat the discussion of
propagators in the field theory contained in~\cite{kos}.} This will
determine the $i\epsilon$ prescription for the bulk-boundary
propagators (\ref{lorp}).

We need to deform the contour to avoid singularities in the
bulk-boundary propagator. The singularities in the
propagator~(\ref{lorp}) are located at
\begin{equation} \label{lsing}
\Delta t = {r_+ r_- \over (r_+^2 - r_-^2)}\Delta \phi_n \pm {r_+ \over
(r_+^2 - r_-^2)} \cosh^{-1} \left[ \sqrt{ \alpha \over \alpha-1} \cosh
r_+ \Delta \phi_n \right] + {2\pi r_+ \over (r_+^2 - r_-^2)} i m,
\end{equation}
where $m \in \mathbb Z$. Note that this has the correct thermal
periodicity in imaginary time.

The structure of the propagator is rather more complicated than in the
non-rotating case. We would therefore first like to check that it has the
correct behaviour as we take the bulk point to large distances---in
this limit, we should recover the singularities of the boundary to
boundary propagator discussed in~\cite{kos}. In the limit $r \to
\infty$, (\ref{lsing}) reduces to
\begin{equation}
\Delta t = { r_+ r_- \pm r_+^2 \over r_+^2 - r_-^2} \Delta \phi_n +
{2\pi r_+ \over r_+^2 - r_-^2} i m.
\end{equation}
If we return to the coordinate $\bar \phi = \phi + r_- t/r_+$ which is
appropriate at large distances, this becomes, after some
rearrangement,
\begin{equation}
\Delta t = \pm \Delta \bar \phi_n + {2\pi r_+ \over (r_+^2 \pm r_+
r_-)} i m,
\end{equation}
reproducing the correct lightcone singularities in the boundary to
boundary propagator.

If we analytically continue $r_- = -i \tilde{r}_-$ the singularities
in the propagator lie at
\begin{equation} \label{esing}
\Delta t = -i {r_+ \tilde{r}_- \over (r_+^2 + \tilde{r}_-^2)} \Delta
\phi_n \pm {r_+ \over (r_+^2 + \tilde{r}_-^2)} \cosh^{-1} \left[
\sqrt{ \alpha \over \alpha-1} \cosh r_+ \Delta \phi_n \right] + {2\pi
r_+ \over (r_+^2 + \tilde{r}_-^2)} i m,
\end{equation}
where $\alpha = (r^2 + \tilde r_-^2)/(r_+^2 + \tilde r_-^2)$. If we
start with the integration over the Euclidean black hole, with the
external source points taken to lie at $t'_i = -i \tau'_i$, then the
appropriate contour is from $t=0$ to $t = -i\beta$. The singularities
in (\ref{esing}) are then of the form imaginary $\pm$ real. That is,
all the singularities associated with future (resp. past) lightlike
separation are to the right (left) of our contour of integration. This
will lead to a standard Feynman $i \epsilon$ prescription when we
rotate back to Lorentzian signature. Note that this would not
necessarily be the case if we tried to define the amplitude via an
integration over imaginary $t$ at real $r_-$. This thus shows the
importance of the analytic continuation in $r_-$ from the point of
view of this calculation.

\begin{figure}
\begin{center}
    \includegraphics[width=\textwidth]{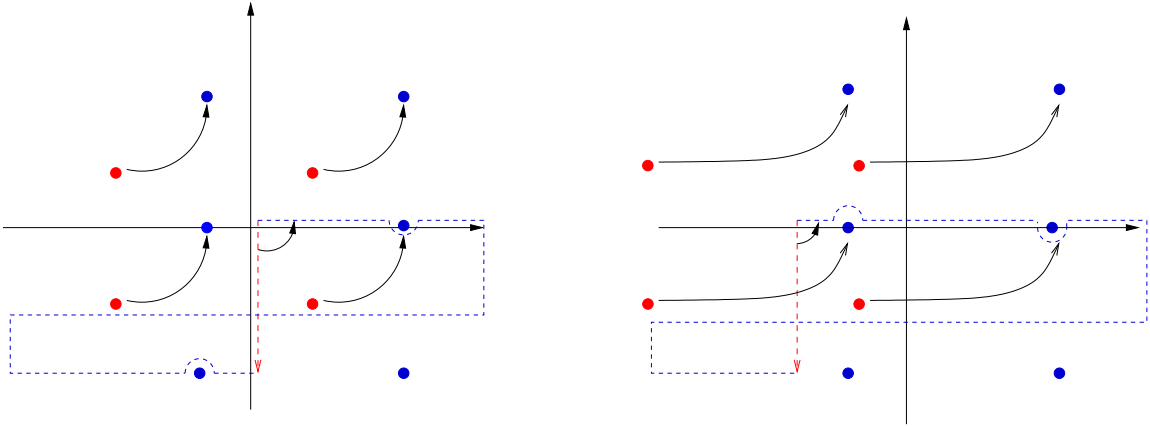}
\end{center}
\caption{The analytic continuation and contour deformation in passing
  from the Euclidean to the Lorentzian section. The complex $z$ plane
  is pictured; on the right, translation invariance has been employed
  to move the original contour of integration to $z \in (-Z, -Z-2\pi
  i)$. To simplify the figure, only singularities associated with an
  external point in region $1_{++}$ have been indicated.}  \label{fig:def}
\end{figure}

Since we rotate $r_-$ as well as $\tau$ in moving from the Euclidean
to the Lorentzian section, it might seem that the motion of the
singularities (and hence the way we need to deform our contour) is
much more complicated in the present case than it was in the
non-rotating case. In fact, it isn't. To follow the motion of the
singularities, it is convenient to define a new variable $z = (r_+^2 -
r_-^2) t/r_+$. Then the locations of the singularities in the complex
$z$ plane are given by
\begin{equation} \label{esingz}
z = z' - \tilde{r}_- i \Delta \phi_n \pm \cosh^{-1} \left[ \sqrt{
\alpha \over \alpha-1} \cosh r_+ \Delta \phi_n \right] + 2\pi i m.
\end{equation}
The only complication then is the behaviour of the third term. But
while we want to continue from the Euclidean section, where $r_- = -i
\tilde r_-$ and $r$ is real, to the Lorentzian section, where $r_-$
and $r$ are real, we need not keep $r$ real at intermediate stages; we
can consider the analytic continuation along any path in the space of
complex metrics that is convenient. Clearly, it is more convenient to
perform the analytic continuation by continuing $z'$ and $r_-$ from
imaginary to real values while keeping $\alpha$ real at intermediate
stages, rather than with $r$ kept real at intermediate stages. The
effect of the analytic continuation will then be to rotate the
singularities onto the real axis by changing the first two terms in
(\ref{esingz}) from imaginary to real values. The rotation moves all
the singularities in a counter-clockwise direction, so if we similarly
move the contour of integration in a counter-clockwise direction,
deforming it as illustrated in figure~\ref{fig:def}, we will obtain
the usual Feynman prescription for integrating over the singularities
on the light-cone.

Thus, after analytic continuation, we can deform the contour of
integration for the vertices to an integration over the real $t$ axis
and $t = -i \pi r_+/(r_+^2-r_-^2)$, and over two vertical segments
connecting the horizontal segments. Since the singularities associated
with external points in region $1_{++}$ will appear on the real axis,
while those associated with external points in region $1_{+-}$ will
appear on the other horizontal segment, we interpret this as saying
that bulk vertices are integrated over the two regions $1_{+\pm}$
outside the black hole's event horizon. The vertical segments create
appropriate correlations in the state on the past and future horizons
on the two sides. This contour also gives us an $i \epsilon$
prescription for the bulk-boundary propagators which goes above the
past singularities and below the future ones on the real axis, and
vice-versa along $t = -i \pi r_+/(r_+^2-r_-^2)$, so it reproduces the
usual thermal propagator in the region outside the black hole on each
side. The addition of non-zero angular momentum therefore does not
introduce any conceptual novelty relative to the discussion for the
non-rotating black hole for the calculation of field theory amplitudes
using this deformation of the contour.

\section{Poincare disc coordinates}
\label{disc}

We would like to relate this calculation of the correlator by an
integration over the region outside the black hole to some integral
which includes the region inside the black hole. To do so, we need to
find a different way of deforming the contour integral as we perform
the analytic continuation. As in~\cite{kos}, the key is to find an
appropriate coordinate system. We want an analogue of the Kruskal
coordinate system used in~\cite{kos}, but now adapted to the
identification used to construct a rotating BTZ black hole. Note that
the Kruskal coordinates introduced for the rotating black hole
in~\cite{btz} are not apt for our purpose, as there is no explicit
expression for the metric as a function of the coordinates in that
coordinate system.

We can construct a coordinate system which will cover the same region
of the spacetime, but in which the metric does have an explicit
expression, by adopting a different perspective on the coordinate
system used in~\cite{kos}. There, the metric on the Euclidean BTZ
black hole was expressed as
\begin{equation} \label{pdisc}
ds^2 = {4 \over (1-\bar X^2 - \bar Y^2)^2} \left[ d\bar Y^2 +
    d\bar X^2 + {r_+^2 \over 4} (1+\bar X^2 + \bar Y^2)^2 d\bar
    \phi^2 \right],
\end{equation}
where $0 \leq \bar X^2 + \bar Y^2 <1$. This coordinate system is
related to the embedding coordinates $(T_1,X_1,X_2,X_3)$ satisfying
$T_1^2-\vec X^2 = 1$ by
\begin{equation}
T_1 = {1+\bar X^2+\bar Y^2 \over 1-\bar X^2-\bar Y^2} \cosh r_+
\bar \phi,
\end{equation}
\begin{equation}
X_1 = {1+\bar X^2+\bar Y^2 \over 1-\bar X^2-\bar Y^2} \sinh r_+
\bar\phi,
\end{equation}
\begin{equation}
X_2 = {2\bar X \over 1-\bar X^2-\bar Y^2}, \quad X_3 = {2\bar Y
  \over 1-\bar X^2-\bar Y^2}.
\end{equation}
We can recognise this coordinate system as a Cartesian version of the
Poincare disc coordinates on (Euclidean) AdS$_3$. That is, the surfaces
$\bar \phi =$ constant in these coordinates are copies of the Poincare
disc, written in a Cartesian coordinate system $\bar X, \bar Y$. If $\bar \phi$ is
unrestricted, this is a familiar description of global AdS$_3$. The
BTZ black hole is obtained by a quotient on hyperbolic space
that is expressed in these coordinates as $\bar \phi \sim \bar \phi +
2\pi$.

Now the difference between the `rotating' and non-rotating Euclidean
black holes is simply that in the former case, we make an
identification with a twist; that is, the identification which
constructs the Euclidean black hole is not orthogonal to the Poincare
discs in this coordinate system. A coordinate system adapted to such
an identification can be created by twisting the above coordinates,
i.e., through
\begin{equation} \label{twist1}
\bar X = X \cos \tilde r_- \phi + Y \sin \tilde r_- \phi,
\end{equation}
\begin{equation} \label{twist2}
\bar Y = -X \sin \tilde r_- \phi + Y \cos \tilde r_- \phi,
\end{equation}
\begin{equation} \label{twist3}
\bar \phi = \phi.
\end{equation}
In these coordinates, the Euclidean metric reads
\begin{eqnarray} \label{ebtzk}
ds^2 &=& {4 \over (1-X^2 -Y^2)^2} \left[ dY^2 +
    dX^2 + 2 \tilde r_- (X dY - Y dX) d\phi + \tilde r_-^2 (X^2 +
    Y^2) d\phi^2 \right. \\ \nonumber && \left. + {r_+^2 \over 4} (1+X^2 +
    Y^2)^2 d\phi^2 \right].
\end{eqnarray}
This coordinate system is related to the embedding coordinates by
\begin{equation}
T_1 = {1+ X^2+ Y^2 \over 1-X^2-Y^2} \cosh r_+
\phi,
\end{equation}
\begin{equation}
X_1 = {1+X^2+Y^2 \over 1-X^2-Y^2} \sinh r_+ \phi,
\end{equation}
\begin{equation}
X_2 = {2 X  \over 1-X^2-Y^2 }\cos \tilde r_- \phi + {2 Y \over
  1-X^2-Y^2 } \sin \tilde r_- \phi,
\end{equation}
\begin{equation}
X_3 = -{2 X \over 1-X^2-Y^2}  \sin \tilde r_- \phi +  {2 Y \over
  1-X^2-Y^2} \cos \tilde r_- \phi.
\end{equation}
The identification $\phi \sim \phi+2\pi$ in these coordinates gives us
the rotating BTZ black hole. Thus, this represents a convenient
coordinate system on the Euclidean black hole.

Let us now analytically continue this coordinate system to Lorentzian
space, setting $T = -iY$ and $r_- = -i \tilde r_-$, and see what
region of the black hole the resulting coordinates cover. We have a
locally AdS$_3$ Lorentzian metric
\begin{eqnarray} \label{btzk}
ds^2 &=& {4 \over (1-X^2 +T^2)^2} \left[ -dT^2 + dX^2 - 2r_- (X dT - T
  dX) d\phi \right. \\ \nonumber && \left. - r_-^2 (X^2 - T^2)
  d\phi^2+ {r_+^2 \over 4} (1+X^2 -T^2)^2 d\phi^2 \right].
\end{eqnarray}
This metric is now related to the Lorentzian embedding coordinates by
\begin{equation}
T_1 = {1+ X^2 -T^2 \over 1-X^2+T^2} \cosh r_+ \phi,
\end{equation}
\begin{equation}
X_1 = {1+X^2-T^2 \over 1-X^2+T^2} \sinh r_+ \phi,
\end{equation}
\begin{equation}
X_2 = {2 X  \over 1-X^2+T^2 }\cosh r_- \phi - {2 T \over
  1-X^2-Y^2 } \sinh r_- \phi,
\end{equation}
\begin{equation}
T_2 = -{2 X \over 1-X^2+T^2}  \sinh r_- \phi +  {2 T \over
  1-X^2+T^2} \cosh r_- \phi.
\end{equation}
Thus, the relation between this coordinate system and the coordinate
system of~(\ref{rbtzb}) is
\begin{equation}
\sqrt{\alpha} = {1+X^2 - T^2 \over 1-X^2 + T^2},
\end{equation}
\begin{equation}
\tanh \left[ {(r_+^2 - r_-^2) \over r_+} t \right] = {T \over X},
\end{equation}
\begin{equation}
\phi = \phi.
\end{equation}
Note that this coordinate transformation is very similar to that
relating the Rindler and Minkowski coordinates on flat space. This
similarity is perhaps more obvious in the inverse coordinate
transformation,
\begin{equation}
X = \left( \sqrt{\alpha} - 1 \over \sqrt{\alpha} + 1 \right)^{1/2}
\cosh  \left( {(r_+^2 - r_-^2) \over r_+} t \right),
\end{equation}
\begin{equation}
T = \left( \sqrt{\alpha} - 1 \over \sqrt{\alpha} + 1 \right)^{1/2}
\sinh  \left( {(r_+^2 - r_-^2) \over r_+} t \right).
\end{equation}

One would like to know what portion of the BTZ metric our rotating
disc coordinates are covering. This is most easily analysed by
following~\cite{kos}, and considering the signs of the combinations
$T_1^2 - X_1^2 = \alpha$ and $T_2^2 - X_2^2 = 1-\alpha$. In the
coordinate system of~(\ref{btzk}), these are
\begin{equation}
T_1^2 - X_1^2 = \left({1+ X^2 -T^2 \over 1-X^2+T^2}\right)^2
\end{equation}
and
\begin{equation}
T_2^2 - X_2^2 = { 4 (T^2 - X^2) \over (1-X^2+T^2)^2}
\end{equation}
We see that $T_1^2 - X_1^2$ is always positive, so we have only
regions with $r > r_-$, while $T_2^2 - X_2^2$ can be positive or
negative, so we have regions with $r>r_+$ or $r<r_+$. Exploring
further, we have
\begin{equation}
T_1 \pm X_1 = {1+ X^2 -T^2 \over 1-X^2+T^2} e^{\pm r_+ \phi}
\end{equation}
\begin{equation}
T_2 \pm X_2 = {2 (T \pm X) \over 1-X^2+T^2} e^{\pm r_- \phi}.
\end{equation}
Thus, we have the following: for $X^2 - T^2 >1$, we have a region
$1_{-+}$ and a region $1_{--}$. The surface $X^2 - T^2 = 1$ maps to
$r=\infty$. For $1> X^2- T^2 \geq 0$, we have a region $1_{++}$ and a
region $1_{+-}$. For $0 \geq X^2-T^2 > -1$, we have a region $2_{++}$
and a region $2_{+-}$. This coordinate system breaks down at $X^2-T^2
= -1$, as the determinant of the metric (\ref{btzk}) vanishes there. For
$-1 > X^2 - T^2 > -\infty$, we have a region $2_{-+}$ and a region
$2_{--}$.

The physical range of coordinates is thus $1 > X^2 - T^2 > -1$, which covers the same region as BTZ Kruskal coordinates $(U_+,V_+)$
\cite{btz}, namely the region including the regions $1_{+\pm}$ outside the horizon and $2_{+\pm}$ between the two horizons in one `copy' of
the BTZ black hole. Just as the regions $1_{-\pm}$ are not physically connected to the regions $1_{+\pm}$, even though they might appear to
be adjacent in these coordinates, the $2_{-\pm}$ are not connected to the regions $2_{+\pm}$ in the physical spacetime (region $3$
intervenes). The presence of these additional regions will be mathematically convenient in the continuation of the amplitudes, but we do not
believe they should be assigned any physical significance.

Let us now consider the propagator in this coordinate system. We know
from the work of~\cite{kos} that the Euclidean propagator on AdS$_3$
expressed in the coordinates~(\ref{pdisc}) is
\begin{equation}
K_E(x,b') = {(1-\bar X^2 - \bar Y^2)^{2h_+} \over [2\bar X \bar X'
    + 2 \bar Y \bar Y' - (1 + \bar X^2 + \bar Y^2) \cosh^2
    r_+ \Delta \bar \phi ]^{2h_+}}.
\end{equation}
Note that since $b = (\bar Y', \bar X',\bar \phi)$ is meant to be a
boundary point, $\bar Y'^2 + \bar X'^2 = 1$.  Using the
coordinate transformation (\ref{twist1}-\ref{twist3}) to our twisted
coordinates, and imposing the identification $\phi \sim \phi+2\pi$, we
find that the Euclidean black hole propagator can be re-expressed in
the coordinates of~(\ref{ebtzk}) as
\begin{eqnarray} \label{ekprop}
K_E(x,b') &=& \sum_{n=-\infty}^{\infty} (1- X^2 - Y^2)^{2h_+} [2 X
  X' \cos \tilde r_- \Delta \phi_n + 2 Y Y' \cos \tilde r_-
  \Delta \phi_n \\ \nonumber &&- 2 X Y' \sin \tilde r_- \Delta
  \phi_n + 2X' Y \sin \tilde r_- \Delta \phi_n - (1 + X^2 + Y^2)
  \cosh r_+ \Delta \phi_n ]^{-2h_+}.
\end{eqnarray}
This can be analytically continued to give us the Lorentzian
propagator in rotating disc coordinates as
\begin{eqnarray} \label{kprop}
K(x,b') &=& \sum_{n=-\infty}^{\infty} (1- X^2 + T^2)^{2h_+} [2 X X'
\cosh r_- \Delta \phi_n - 2 T T' \cosh r_- \Delta
\phi_n \\ \nonumber && + 2 X T' \sinh r_- \Delta \phi_n - 2X' T \sinh r_-
\Delta \phi_n - (1 + X^2 - T^2) \cosh r_+ \Delta \phi_n ]^{-2h_+}.
\end{eqnarray}
Before working out the singularities in this propagator in detail, let
us note an important change in the singularity structure. In the
non-rotating case, there was a divergence of the summation at $X^2 -
T^2 = -1$, which in that case corresponds to the location of the BTZ
singularity. This reflected the fact that as we approach the BTZ
singularity, the proper distance between different images of the bulk
point in the covering space is going to zero (as is obvious from the
fact that $g_{\phi \phi}$ is vanishing there). For non-zero rotation,
there is no analogue of this effect, because even for $X^2-T^2=-1$,
the summand does not become independent of $n$. This was already seen
in BTZ coordinates in (\ref{lorpch}).

This is physically correct; in the non-rotating case, $X^2-T^2=-1$ is
the BTZ singularity, and as was pointed out in~\cite{exciting,eskos,kos},
the BTZ singularity is like the singularity in the parabolic orbifold
of flat space studied in~\cite{liu,fabinger,lawrence,hp}. The
divergence in the summation in the propagator near the BTZ singularity
is thus related to the instability of this singularity identified
in~\cite{hp} (where multiple boosted images of a single particle
converge).  In the rotating black hole, by contrast, $g_{\phi \phi}$
does not vanish at $X^2 - T^2 = -1$, so the proper distance between
images remains finite at $X^2- T^2=-1$; this is just a horizon in the
spacetime. As discussed in the introduction, we nevertheless expect
that there will be a singularity there, due to the infinite blueshift
associated with this horizon, so the cutoff at the event horizon is
not just an artifact of our coordinate system. The curvature
singularity that forms at the Cauchy Horizon is a weak, null
singularity (which is characterised by having finite tidal
deformations)~\cite{sing1,sing2}. As a result the
singularity is neither associated with the sum over images or the
divergence of the probe field. Hence, we should not expect it to be
signalled by a failure of the summation in (\ref{kprop}) to converge
or the bulk-boundary propagator to in fact diverge at all.

Equally striking is what adding rotation does not change: there is
still no imaginary periodicity in $T$. The singularities in the
propagator are given by solving a quadratic in $T$, so there will be
singularities at two locations in the complex plane for given $X, X',
T', \Delta \phi_n$. Thus, perhaps surprisingly, there is no thermal
behaviour associated with the inner horizon in this representation of
the propagator, even though we are restricted to the region outside
the inner horizon. We would argue that this is again a sign that the
inner horizon is not a true regular horizon in the spacetime, because
of the blueshifting instability.

Let us now determine the singularities in this Lorentzian propagator
in detail. Note that the source point is by assumption located on the
boundary; we established earlier that this corresponds to $X'^2- T'^2
= 1$. It is useful to introduce a new variable $\beta$ parameterising
the location of the source point, writing $X' = \eta \cosh \beta$, $T'
= \eta \sinh \beta$, where $\eta=\pm 1$. The singularities are then
given by the roots of the quadratic in~(\ref{kprop}):
\begin{eqnarray}
2 X \eta \cosh \beta
\cosh r_- \Delta \phi_n - 2 T \eta \sinh \beta \cosh r_- \Delta
\phi_n + 2 X \eta \sinh \beta  \sinh r_- \Delta \phi_n&& \\  \nonumber - 2 T
\eta \cosh \beta \sinh r_-
\Delta \phi_n - (1 + X^2 - T^2) \cosh r_+ \Delta \phi_n &=& 0,
\end{eqnarray}
which we can simplify to
\begin{equation}
T^2 \cosh r_+ \Delta \phi_n - 2 \eta T \sinh (\beta - r_- \Delta \phi_n)
- (1 + X^2) \cosh r_+ \Delta \phi_n +2 \eta X \cosh(\beta - r_- \Delta
\phi_n) =0.
\end{equation}
The roots lie at
\begin{eqnarray}
T&=& (\cosh r_+ \Delta \phi_n)^{-1} \left\{ \eta \sinh(\beta - r_-
\Delta \phi_n) \right. \\ && \left. \pm [\sinh^2(\beta - r_- \Delta
\phi_n) + (1 + X^2) \cosh^2 r_+ \Delta \phi_n - 2 \eta X \cosh(\beta - r_-
\Delta \phi_n)\cosh r_+ \Delta \phi_n]^{1/2}
\right\}. \nonumber
\end{eqnarray}
This can be further simplified to read
\begin{equation} \label{eksing}
T =  {  \eta \sinh(\beta - r_- \Delta \phi_n) \pm [( \cosh (\beta - r_-
\Delta \phi_n) - \eta X \cosh r_+ \Delta \phi_n)^2 + \sinh^2
r_+ \Delta \phi_n]^{1/2}
\over    \cosh r_+ \Delta \phi_n}.
\end{equation}
Note that the two singularities in the complex $T$ plane lie on the
real axis, as they should; these are just light-cone singularities.

We would now like to study the contour deformation for the vertex
using this coordinate system, and thereby obtain an $i \epsilon$
prescription for the Lorentzian propagator (\ref{kprop}). As
in~\cite{kos}, there is a technical difficulty in that the natural
range of integration in the Euclidean metric is $0 \leq X^2 + Y^2
\leq 1$, and we do not understand how to perform an analytic
continuation subject to this kind of restriction. However, the
propagator (\ref{ekprop}) is symmetric under the antipodal map
\begin{equation}
X \to {X \over X^2 + Y^2}, \quad Y \to {Y \over X^2 + Y^2}
\end{equation}
so long as $2h_+$ is even. Hence, at least in this case, we can treat
the region $X^2 + Y^2 > 1$ as another copy of the black hole
solution, and obtain the correct answer for the amplitude by
integrating over the full $X,Y$ plane and then dividing the final
answer by two.

Let us therefore consider an integration over the Euclidean section,
where the source points lie on the circle $X'^2 + Y'^2 = 1$, and
the integration over $T$ runs from $T = i\infty$ to $T= -i\infty$. The
singularities then lie initially at
\begin{equation}
T = { -i\eta \sin(\tilde \beta - \tilde r_- \Delta \phi_n) \pm [( \cos
    (\tilde \beta - \tilde r_- \Delta \phi_n) - \eta X \cosh r_+
    \Delta \phi_n)^2 + \sinh^2 r_+ \Delta \phi_n]^{1/2} \over \cosh
  r_+ \Delta \phi_n},
\end{equation}
where $\tilde \beta$ is defined by $X' = \eta \cos \tilde \beta$,
$Y' = \eta \sin \tilde \beta$. As in the BTZ coordinates,
we see that the singularities associated with the Euclidean solution
lie at imaginary $\pm$ real, so that the singularities associated with
future (resp. past) lightlike separation are to the right (left) of
our contour of integration.

\begin{figure}
\begin{center}
   \includegraphics[width=0.6\textwidth]{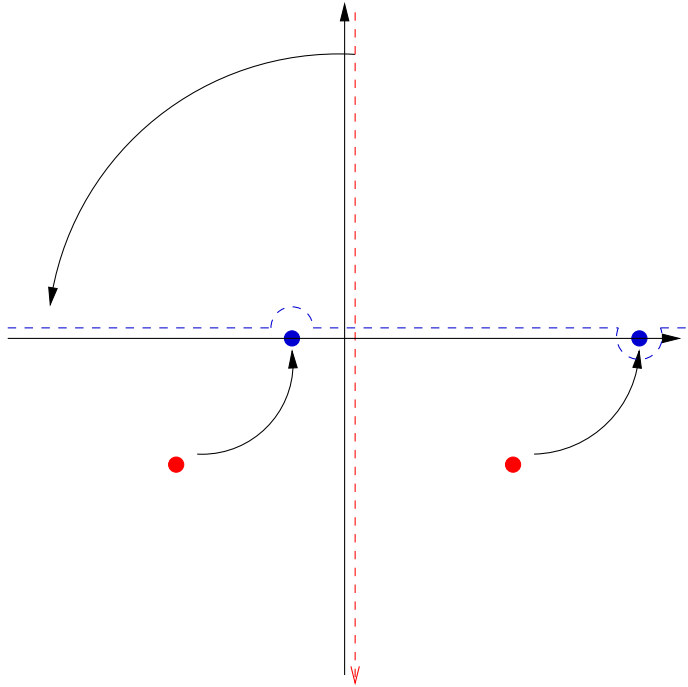}
\end{center}
\caption{The analytic continuation and contour deformation in passing
  from the Euclidean to the Lorentzian section, in Poincare disc
  coordinates.} \label{fig:def2}
\end{figure}

The analytic continuation here looks still more complicated than in
the previous case; however, once again we can avoid complications in
the motion of the singularities by a convenient choice of
path. Instead of keeping $X$ real at intermediate stages in the
analytic continuation, it is better to assume we keep $\cosh ( \beta -
r_- \Delta \phi_n) - \eta X \cosh r_+ \Delta \phi_n$ real. The
analytic continuation then involves only the first term in
(\ref{eksing}), and will rotate the singularities
counter-clockwise. If we similarly rotate the contour
counter-clockwise, we will wind up with the usual $i \epsilon$
prescription, where we make the substitution $Y = e^{i(\pi/2 -
  \epsilon)} T$ in (\ref{ekprop}). The analytic continuation is thus
accurately represented by the simple picture given in figure~\ref{fig:def2}.

We are left with an integration over the full $X, T$ plane for each
vertex. However, as we saw above, the full $X, T$ plane contains
regions $1_{-\pm}$ and $2_{-\pm}$, as well as the regions $1_{+\pm},
2_{+\pm}$ of interest. It is thus a double cover of a single Kruskal
patch $r \geq r_-$ in the Lorentzian black hole. We can simply cancel
this double cover against the factor of a half that came from the
double cover we introduced in the Euclidean black hole, and obtain an
integration just over the region $-1 \leq X^2 - T^2 < 1$
corresponding to a single copy of the Kruskal patch. As in~\cite{kos},
this will lead to a slightly more complicated $i \epsilon$
prescription in the regions $r_- \leq r \leq r_+$.

The main point is that, by analytic continuation, we can express the
bulk spacetime version of the calculation of a CFT amplitude either
in terms of an integration of the interaction vertices over regions
$1_{++}$ and $1_{+-}$ in the rotating black hole, with the $i
\epsilon$ prescription given in the previous section, or in terms of
an integration over regions $1_{+\pm}$ and $2_{+\pm}$, with the $i
\epsilon$ prescription given here. Thus, such amplitudes involve the
region behind the event horizon, but outside of the Cauchy horizon.

\section{Conclusions}
\label{concl}

Our main objective in this paper has been to deepen our understanding
of the relation between the field theory and the spacetime in the
AdS/CFT correspondence for spacetimes with horizons. We have shown
that in rotating BTZ, the bulk calculation of field theory amplitudes
defined by analytic continuation can be expressed in two ways, in
terms of an integral over the region outside the horizon, or in terms
of an integral over the region bounded by past and future Cauchy
horizons. This extends previous work~\cite{kos} to include
rotation.

A feature which deserves to be stressed again is that the analytic
continuation from the Euclidean black hole gives us an integral over
both asymptotic regions, $1_{++}$ and $1_{+-}$. In particular, the
analytic continuation of the single boundary in Euclidean BTZ gives us
the two boundaries in Lorentzian BTZ. In the original coordinate
system, this arises because the deformed contour defining the
real-time thermal field theory involves two horizontal segments,
interpreted as independent fields on the two boundaries that are
entangled by the state of the CFT. In the Poincare disc coordinates,
we pass from the connected boundary $X^2 + Y^2 = 1$ in Euclidean
space to the two-sheeted boundary $X^2 - T^2 = 1$ in Lorentzian
space. The black hole is thus still represented by a pure state---an
entangled state of the two field theories. It seems natural that such
a pure state will contain complete information about the bulk
spacetime. If we trace over the degrees of freedom on one boundary,
the resulting thermal state on the other boundary will no longer
contain enough information to describe the region behind the horizon.

The main value of adding rotation is that it highlights the importance
of the Cauchy horizon as a boundary. We have seen that amplitudes can
be related to an integral over the region bounded by the Cauchy
horizons. For Lorentzian AdS, a state in the field theory is
conjectured~\cite{hads} to be dual to the initial state in the bulk
AdS spacetime, encoded by data on some Cauchy surface. It therefore
seems very natural that the calculation of amplitudes in the field
theory involves the region determined by this initial data. We
conjecture that in general, the CFT description is not limited to the
spacetime region which is visible from infinity---that is, it does not
stop at event horizons. It stops only when we encounter a Cauchy
horizon.\footnote{A similar argument that the Lorentzian
correspondence requires that some regions beyond event horizons be
included in the CFT description was made in~\cite{hub}.}

This conjecture gains further support from carefully considering the
horizon in the Poincare coordinates on pure AdS. In~\cite{gh}, it was
found that the region behind the horizon in Poincare coordinates is
included in the CFT description, because the CFT naturally lives on the
whole Einstein static universe $\mathbb R \times S^n$, and not just in
Minkowski space $\mathbb R^{n,1}$. Put another way, the CFT on
$\mathbb R^{n,1}$ determines the bulk spacetime within the Poincare
horizon; to go beyond it, we need to fix boundary conditions on other
parts of the boundary of AdS. The boundary of the Poincare patch is an
(observer-dependent) event horizon, but it is also the Cauchy horizon
for a $t=$ constant surface in Poincare coordinates (supplemented by
the boundary conditions supplied by the field theory on $\mathbb
R^{n,1}$). Thus, the result of~\cite{gh}, that an extension of the
field theory (from $\mathbb R^{n,1}$ to $\mathbb R \times S^n$) is
necessary to describe the region of spacetime beyond the Poincare
horizon, is perfectly consonant with relating the field theory to the
region bounded by the Cauchy horizons.

In fact, our consideration of the rotating BTZ black hole is very
close in spirit to the analysis of~\cite{gh}. The periodic
identification in $\phi$ that defines the black hole has played only a
spectator role in our discussion. Thus, one could equally well apply
our analysis to AdS in BTZ coordinates without the identification
along $\phi$. We would then say that what we have learnt is that if
one considers the the field theory just on those regions of the
boundary where $\partial_\phi$ is spacelike, it is dually related to
the region of the spacetime which is determined by Cauchy data
supplemented by the asymptotic boundary conditions on this portion of
the boundary. So the discussion here is in a sense a generalisation
of~\cite{gh}, considering the restriction to a different region of the
boundary.

If the field theory only describes the region of spacetime before the
Cauchy horizon, what is the physical interpretation of the region
beyond that horizon? In some special cases, such as the Poincare
coordinates in pure AdS, this is an unnatural restriction, and the
spacetime beyond the horizon can be included in the description by an
extension of the field theory which supplies the necessary boundary
conditions. However, we will further conjecture that generically there
will be no such extension---the region beyond the Cauchy horizon is
not included in the `hologram', so it is unphysical. Thus, the AdS/CFT
correspondence enforces strong cosmic censorship for such
solutions~\cite{scc}. The classical description will break down at the
Cauchy horizon, so no observer can study quantum effects from a safe
distance.

The idea that the spacetime is cut off at the Cauchy horizon receives
support from previous calculations, which have shown that an observer
approaching the Cauchy horizon inside a rotating black hole will
generically encounter a divergent flux of radiation due to the
infinite blueshift~\cite{instab}. Thus, the classical theory does
predict its own downfall there. This should be contrasted with the
situation at the event horizon, where an observer crossing the horizon
will see nothing special. An important direction for future work is to
understand to what extent this blueshifting instability is reflected
in the CFT calculations we have discussed here, and more generally to
see evidence for the conjectured breakdown of the classical spacetime
description at the Cauchy horizon and understand what replaces it.

If the event horizon is not a fundamental barrier, why does it remain
so difficult to answer questions about physics there? One part of the
answer to this question, as pointed out in~\cite{kos}, is that the
size of a black hole, from the viewpoint of an infalling observer, is
never greater than the characteristic curvature scale of the AdS
space. This remains true for rotating black holes: the proper time
taken to fall from the outer to the inner event horizon of a rotating
black hole is bounded from above by a number of order one in AdS units
for any value of $r_\pm$. Hence, one obstacle to understanding the
observations of infalling observers is that we do not know how to
recover approximately local bulk physics on scales smaller than the AdS
scale from the field theory. Making progress on this issue will be
essential to gain a better understanding of the description of the
black hole's geometry in the field theory.

Another barrier to understanding is that the part of the connection
between bulk and boundary we understand well is the relation of the
asymptotic behaviour of fields in spacetime and the expectation values
of local operators in the field theory. This can be used indirectly to
understand some features of the description of sources in the bulk in
terms of their effects at large distances, but does not tell us how
the CFT describes changes in the bulk whose effects have not yet
propagated to the boundary. Understanding the CFT degrees of freedom
which describe the bulk more directly is the problem of
precursors~\cite{pre1,pre2,pre3,pre4}. Precursors are supposed to
reflect events deep inside the spacetime without (immediately)
changing any local observables in the CFT. The region behind the black
hole's event horizon provides a particularly sharp example of this
problem, as events in this region can never effect the asymptotic
region, so they will only ever be reflected in the CFT through
non-local observables tied to the precursors. This point was also
discussed in~\cite{hub}. Thus, understanding the description of
changes behind the event horizon will also require progress on our
understanding of the direct description of the interior of spacetime
in the dual CFT. The present calculation reinforces the arguments that the
region inside the event horizon is included in the CFT description,
but it does not address these deep questions concerning how that
description is implemented.

Finally, it would be interesting to explore the extension of these
calculations to higher dimensions. This will clearly be more
challenging, as the construction of suitable coordinate systems in our
work here has relied on cunningly exploiting the locally AdS character
of the BTZ black holes.

\centerline{\bf Acknowledgements}
\medskip

We thank Vijay Balasubramanian and Asad Naqvi for extensive
discussions. T.S.L. is supported by the DOE under cooperative research
agreement DE-FG02-95ER40893. S.F.R. is supported by an EPSRC Advanced
Fellowship.


\begin{thebibliography}{19}

\bibitem{kos}
P.~Kraus, H.~Ooguri and S.~Shenker,
``Inside the horizon with AdS/CFT,''
arXiv:hep-th/0212277.
%%CITATION = HEP-TH 0212277;%%

\bibitem{dual1}
J.~M.~Maldacena,
``The large $N$ limit of superconformal field theories and supergravity,''
Adv.\ Theor.\ Math.\ Phys.\  {\bf 2} (1998) 231
[Int.\ J.\ Theor.\ Phys.\  {\bf 38} (1999) 1113]
[arXiv:hep-th/9711200].
%%CITATION = HEP-TH 9711200;%%

\bibitem{dualbh}
E.~Witten,
``Anti-de Sitter space, thermal phase transition, and confinement in  gauge theories,''
Adv.\ Theor.\ Math.\ Phys.\  {\bf 2} (1998) 505
[arXiv:hep-th/9803131].
%%CITATION = HEP-TH 9803131;%%

\bibitem{resol}
G.~T.~Horowitz and S.~F.~Ross,
``Possible resolution of black hole singularities from large $N$ gauge  theory,''
JHEP {\bf 9804} (1998) 015
[arXiv:hep-th/9803085].
%%CITATION = HEP-TH 9803085;%%

\bibitem{compa}
L.~Susskind and L.~Thorlacius,
``Gedanken experiments involving black holes,''
Phys.\ Rev.\ D {\bf 49} (1994) 966
[arXiv:hep-th/9308100].
%%CITATION = HEP-TH 9308100;%%

\bibitem{compb}
L.~Susskind, L.~Thorlacius and J.~Uglum,
``The Stretched horizon and black hole complementarity,''
Phys.\ Rev.\ D {\bf 48}, 3743 (1993)
[arXiv:hep-th/9306069].
%%CITATION = HEP-TH 9306069;%%

\bibitem{holog}
V.~Balasubramanian and S.~F.~Ross,
``Holographic particle detection,''
Phys.\ Rev.\ D {\bf 61} (2000) 044007
[arXiv:hep-th/9906226].
%%CITATION = HEP-TH 9906226;%%

\bibitem{geon}
J.~Louko, D.~Marolf and S.~F.~Ross,
``On geodesic propagators and black hole holography,''
Phys.\ Rev.\ D {\bf 62} (2000) 044041
[arXiv:hep-th/0002111].
%%CITATION = HEP-TH 0002111;%%

\bibitem{eternal}
J.~M.~Maldacena,
``Eternal black holes in Anti-de-Sitter,''
arXiv:hep-th/0106112.
%%CITATION = HEP-TH 0106112;%%

\bibitem{hub}
V.~E.~Hubeny,
``Precursors see inside black holes,''
arXiv:hep-th/0208047.
%%CITATION = HEP-TH 0208047;%%

\bibitem{thermof}
W.~Israel,
``Thermo Field Dynamics Of Black Holes,''
Phys.\ Lett.\ A {\bf 57} (1976) 107.
%%CITATION = PHLTA,A57,107;%%

\bibitem{orb1}
J.~M.~Maldacena and A.~Strominger,
``AdS$_3$ black holes and a stringy exclusion principle,''
JHEP {\bf 9812} (1998) 005
[arXiv:hep-th/9804085].
%%CITATION = HEP-TH 9804085;%%

\bibitem{orb2}
G.~T.~Horowitz and D.~Marolf,
``A new approach to string cosmology,''
JHEP {\bf 9807} (1998) 014
[arXiv:hep-th/9805207].
%%CITATION = HEP-TH 9805207;%%

\bibitem{hads}
V.~Balasubramanian, P.~Kraus and A.~E.~Lawrence,
``Bulk vs. boundary dynamics in anti-de Sitter spacetime,''
Phys.\ Rev.\ D {\bf 59} (1999) 046003
[arXiv:hep-th/9805171].
%%CITATION = HEP-TH 9805171;%%
V.~Balasubramanian, P.~Kraus, A.~E.~Lawrence and S.~P.~Trivedi,
``Holographic probes of anti-de Sitter space-times,''
Phys.\ Rev.\ D {\bf 59} (1999) 104021
[arXiv:hep-th/9808017].
%%CITATION = HEP-TH 9808017;%%

\bibitem{berk}
M.~Berkooz, B.~Craps, D.~Kutasov and G.~Rajesh,
``Comments on cosmological singularities in string theory,''
JHEP {\bf 0303} (2003) 031
[arXiv:hep-th/0212215].
%%CITATION = HEP-TH 0212215;%%

\bibitem{scc}
R.~Penrose,
``Gravitational Collapse: The Role Of General Relativity,''
Riv.\ Nuovo Cim.\  {\bf 1} (1969) 252
[Gen.\ Rel.\ Grav.\  {\bf 34} (2002) 1141];
%%CITATION = RNCIB,1,252;%%
R.~Penrose, ``Singularities and time-asymmetry,'' in {\it General
Relativity, an Einstein Centenary Survey}, eds. S.W.~Hawking and
W.~Israel (Cambridge University Press, 1979).

\bibitem{blues}
R.~Penrose, in {\it Batelle Rencontres}, eds. C.~de Witt and
J.~Wheeler (W.A. Benjamin, New York, 1968), p. 222.

\bibitem{instab}
R.~A.~Matzner, N.~Zamorano and V.~D.~Sanberg,
``Instability of the Cauchy horizon of Reissner-Nordstr\"om black holes,''
Phys.\ Rev.\  D {\bf 19}, 2821 (1979);
%%CITATION = PHRVA,D19,2821;%%
S.~Chandrasekhar and J.~Hartle, ``On crossing the Cauchy horizon of a
Reissner-Nordstr\"om black hole,''
Proc.\ Roy.\ Soc.\ Lond.\  {\bf A384}, 301 (1982).
%%CITATION = PRSLA,A384,301;%%

\bibitem{btz}
M.~Banados, M.~Henneaux, C.~Teitelboim and J.~Zanelli,
``Geometry of the (2+1) black hole,''
Phys.\ Rev.\ D {\bf 48}, 1506 (1993)
[arXiv:gr-qc/9302012].
%%CITATION = GR-QC 9302012;%%

\bibitem{eskos}
S.~Hemming, E.~Keski-Vakkuri and P.~Kraus,
``Strings in the extended BTZ spacetime,''
JHEP {\bf 0210} (2002) 006
[arXiv:hep-th/0208003].
%%CITATION = HEP-TH 0208003;%%

\bibitem{esko}
E.~Keski-Vakkuri,
``Bulk and boundary dynamics in BTZ black holes,''
Phys.\ Rev.\ D {\bf 59}, 104001 (1999)
[arXiv:hep-th/9808037].
%%CITATION = HEP-TH 9808037;%%

\bibitem{exciting}
E.~J.~Martinec and W.~McElgin,
``Exciting AdS orbifolds,''
JHEP {\bf 0210} (2002) 050
[arXiv:hep-th/0206175].
%%CITATION = HEP-TH 0206175;%%

\bibitem{liu}
H.~Liu, G.~Moore and N.~Seiberg,
``Strings in a time-dependent orbifold,''
JHEP {\bf 0206} (2002) 045
[arXiv:hep-th/0204168];
%%CITATION = HEP-TH 0204168;%%
H.~Liu, G.~Moore and N.~Seiberg,
``Strings in time-dependent orbifolds,''
JHEP {\bf 0210} (2002) 031
[arXiv:hep-th/0206182].
%%CITATION = HEP-TH 0206182;%%

\bibitem{lawrence}
A.~Lawrence,
``On the instability of 3D null singularities,''
JHEP {\bf 0211} (2002) 019
[arXiv:hep-th/0205288].
%%CITATION = HEP-TH 0205288;%%

\bibitem{fabinger}
M.~Fabinger and J.~McGreevy,
``On smooth time-dependent orbifolds and null singularities,''
arXiv:hep-th/0206196.
%%CITATION = HEP-TH 0206196;%%

\bibitem{hp}
G.~T.~Horowitz and J.~Polchinski,
``Instability of spacelike and null orbifold singularities,''
Phys.\ Rev.\ D {\bf 66} (2002) 103512
[arXiv:hep-th/0206228].
%%CITATION = HEP-TH 0206228;%%

\bibitem{sing1}
E.~Poisson and W.~Israel,
``Internal Structure Of Black Holes,''
Phys.\ Rev.\ D {\bf 41} (1990) 1796.
%%CITATION = PHRVA,D41,1796;%%

\bibitem{sing2}
A.~Ori,
``Inner structure of a charged black hole: An exact mass-inflation solution''
Phys.\ Rev.\ Lett.\  {\bf 67} (1991) 789.
%%CITATION = PRLTA,67,789;%%

\bibitem{gh}
G.~T.~Horowitz and H.~Ooguri,
``Spectrum of large N gauge theory from supergravity,''
Phys.\ Rev.\ Lett.\  {\bf 80} (1998) 4116
[arXiv:hep-th/9802116].
%%CITATION = HEP-TH 9802116;%%

\bibitem{pre1}
J.~Polchinski, L.~Susskind and N.~Toumbas,
``Negative energy, superluminosity and holography,''
Phys.\ Rev.\ D {\bf 60} (1999) 084006
[arXiv:hep-th/9903228].
%%CITATION = HEP-TH 9903228;%%

\bibitem{pre2}
L.~Susskind and N.~Toumbas,
``Wilson loops as precursors,''
Phys.\ Rev.\ D {\bf 61} (2000) 044001
[arXiv:hep-th/9909013].
%%CITATION = HEP-TH 9909013;%%

\bibitem{pre3}
S.~B.~Giddings and M.~Lippert,
``Precursors, black holes, and a locality bound,''
Phys.\ Rev.\ D {\bf 65} (2002) 024006
[arXiv:hep-th/0103231].
%%CITATION = HEP-TH 0103231;%%

\bibitem{pre4}
B.~Freivogel, S.~B.~Giddings and M.~Lippert,
``Toward a theory of precursors,''
Phys.\ Rev.\ D {\bf 66} (2002) 106002
[arXiv:hep-th/0207083].
%%CITATION = HEP-TH 0207083;%%


\end{thebibliography}
\end{document}